\documentclass[%
 reprint,
 amsmath,amssymb,
 aps,
 prc,
floatfix,
]{revtex4-1}

\usepackage{graphicx}% Include figure files
\usepackage{dcolumn}% Align table columns on decimal point
\usepackage{bm}% bold math
\usepackage{float}
\begin{document}

\preprint{APS/123-QED}

\title{Investigation of the reaction $^{74}$Ge(p,$\gamma$)$^{75}$As using the in-beam method
 to improve reaction network predictions for $p$ nuclei}

\author{A.~Sauerwein}
\email[]{sauerwein@ikp.uni-koeln.de}
\affiliation{Institut f\"ur Kernphysik, Universit\"at zu K\"oln, Z\"ulpicher Stra\ss{}e 77, 50937 K\"oln, Germany}
\author{V.~Foteinou}
\affiliation{Tandem Accelerator Laboratory, Institute of Nuclear Physics, NCSR ``Demokritos``, 153.10 Aghia Paraskevi, Athens, Greece}
\author{G.~Provatas}
\affiliation{Tandem Accelerator Laboratory, Institute of Nuclear Physics, NCSR ``Demokritos``, 153.10 Aghia Paraskevi, Athens, Greece}
\author{T.~Konstantinopoulos}
\affiliation{Tandem Accelerator Laboratory, Institute of Nuclear Physics, NCSR ``Demokritos``, 153.10 Aghia Paraskevi, Athens, Greece}
\author{M.~Axiotis}
\affiliation{Tandem Accelerator Laboratory, Institute of Nuclear Physics, NCSR ``Demokritos``, 153.10 Aghia Paraskevi, Athens, Greece}
\author{S. F.~Ashley}
\affiliation{Tandem Accelerator Laboratory, Institute of Nuclear Physics, NCSR ``Demokritos``, 153.10 Aghia Paraskevi, Athens, Greece}
\author{S.~Harissopulos}
\affiliation{Tandem Accelerator Laboratory, Institute of Nuclear Physics, NCSR ``Demokritos``, 153.10 Aghia Paraskevi, Athens, Greece}
\author{J.~Endres}
\affiliation{Institut f\"ur Kernphysik, Universit\"at zu K\"oln, Z\"ulpicher Stra\ss{}e 77, 50937 K\"oln, Germany}
\author{L.~Netterdon}
\affiliation{Institut f\"ur Kernphysik, Universit\"at zu K\"oln, Z\"ulpicher Stra\ss{}e 77, 50937 K\"oln, Germany}
\author{T.~Rauscher}
\affiliation{Department of Physics, University of Basel, Klingelbergstra\ss{}e 82, 4056 Basel, Switzerland}
\author{A.~Zilges}
\affiliation{Institut f\"ur Kernphysik, Universit\"at zu K\"oln, Z\"ulpicher Stra\ss{}e 77, 50937 K\"oln, Germany}

\date{\today}% It is always \today, today,

\begin{abstract}
\begin{description}
\item[Background]
Astrophysical models studying the origin of the neutron-deficient 
$p$ nuclides require the knowledge of proton capture cross sections 
at low energy. The production site of the $p$ nuclei is still under 
discussion but a firm basis of nuclear reaction rates is required 
to address the astrophysical uncertainties. Data at astrophysically 
relevant interaction energies are scarce. Problems with the prediction 
of charged particle capture cross sections at low energy were 
found in the comparisons between previous data and calculations in 
the Hauser-Feshbach statistical model of compound reactions.
\item[Purpose]
A measurement of $^{74}$Ge(p,$\gamma$)$^{75}$As at low proton 
energies, inside the astrophysically relevant energy region, is 
important in several respects. The reaction is directly important as
it is a bottleneck in the reaction flow which produces the lightest $p$ 
nucleus $^{74}$Se. It is also an important addition to the 
data set required to test reaction-rate predictions and to allow an 
improvement in the global p+nucleus optical potential required in 
such calculations. 
\item[Method]
An in-beam experiment was performed, making it possible to measure in the 
range $2.1 \leq E_{p}\leq 3.7~\mathrm{MeV}$, which is for the most 
part inside the astrophysically relevant energy window. Angular distributions 
of the $\gamma$-ray transitions were measured with high-purity germanium detectors at eight 
angles relative to the beam axis. In addition to the total cross sections, 
partial cross sections for the direct population of twelve levels were 
determined.
\item[Results]
The resulting cross sections were compared to Hauser-Feshbach calculations 
using the code SMARAGD. Only a constant renormalization factor of the 
calculated proton widths allowed a good reproduction of both total 
and partial cross sections. The accuracy of the calculation made it possible
to check the spin assignment of some states in $^{75}$As. In the case
of the 1075 keV state, a double state with spins and parities of 3/2$^-$
and 5/2$^-$ is needed to explain the experimental partial cross sections. 
A change in parity from 5/2$^+$ to 5/2$^-$ is required for the
state at 401 keV. Furthermore, in the case of $^{74}$Ge, studying the combination
of total and partial cross sections made it possible to test the $\gamma$ width, which is 
essential in the calculation of the astrophysical $^{74}$As(n,$\gamma$)$^{75}$As 
rate.
\item[Conclusions]
Between data and statistical model prediction a factor of about two was found.
Nevertheless, the improved 
astrophysical reaction rate of $^{74}$Ge(p,$\gamma$) (and its reverse 
reaction) is only 28\% larger than the previous standard rate. The 
prediction of the $^{74}$As(n,$\gamma$)$^{75}$As rate (and its reverse) 
was confirmed, the newly calculated rate differs only by a few 
percent from the previous prediction. The in-beam method with high 
efficiency detectors proved to be a powerful tool for studies in nuclear 
astrophysics and nuclear structure.
\end{description}
\end{abstract}

\pacs{Valid PACS appear here}% PACS, the Physics and Astronomy
\maketitle

%\tableofcontents

 \section{\label{sec:introduction}Introduction}

 \begin{figure*}
 \includegraphics[width=\textwidth]{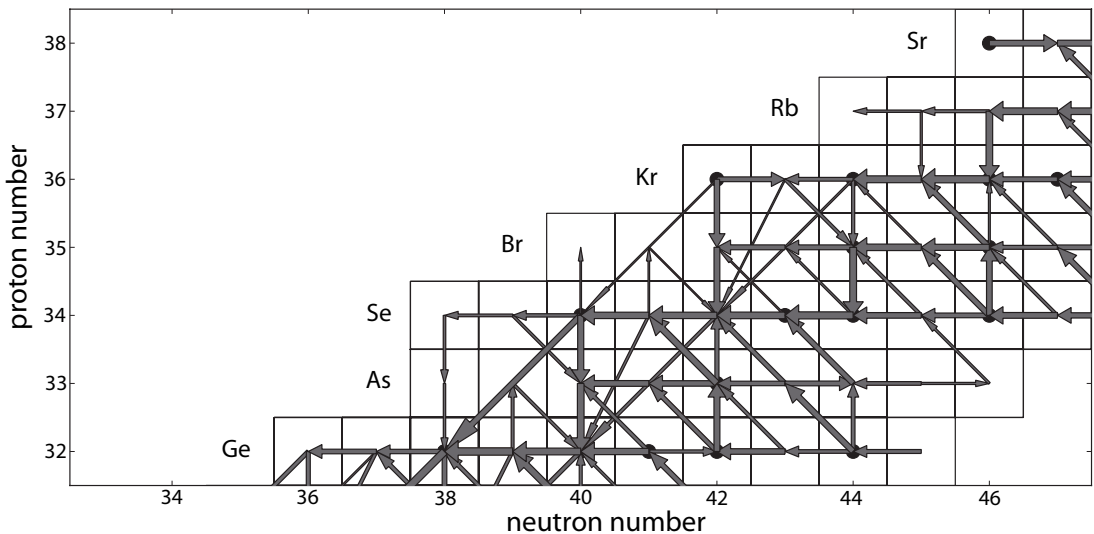}
 \caption{\label{fig:flow} Reaction flow $f$ in the $\gamma$ process 
 in the vicinity of the $p$ nucleus $^{74}$Se. Depicted are 
 time-integrated fluxes from the layer of a 25 $M_\odot$ star which 
 effectively produces $^{74}$Se when the shockfront of a core-collapse 
 supernova explosion is passing through. Arrow thicknesses give the 
 magnitude of the flow on a logarithmic scale.}
 \end{figure*}

About 35 proton-rich nuclei \cite{Arlandini,Nemeth,Arnould}, the so-called 
$p$ nuclei, cannot be synthesized by neutron capture reactions via the 
$s$ and $r$ process \cite{Burbidge,Cameron,Cowan,kapgall,AGT}. The actual astrophysical 
environment for the production of the $p$ nuclei is still under discussion. 
The long-time favored site, explosive burning in the O/Ne shell of massive 
stars before and during a supernova explosion, was found to 
underproduce significantly the light $p$ nuclei with $A<100$ 
\cite{Woosley78,rayet95,rhhw02,Arnould,hegXX}. An alternative would be that 
not just a single process is responsible for all $p$ nuclei but that 
different processes in a number of astrophysical sites produce certain 
ranges of $p$ nuclei (see \cite{Rauscher_NIC} and references therein).
%In principle, there are two ways to produce proton-rich nuclides. First, 
%by adding protons to lighter nuclei via proton-capture reactions or 
%secondly by removing neutrons from heavier seed nuclei by sequences of 
%($\gamma$,x) photo-disintegration reactions. Adding protons is only 
%feasible for the light $p$ nuclei under extremely proton-rich 
%conditions, whereas photodisintegration reactions
%require a finely tuned range of sufficiently high temperatures, 
%which are only achievable  in explosive nuclear burning.
In the search for the $p$-nucleus production mechanism a number of 
processes were suggested, such as the $\gamma$ process in massive stars 
\cite{Woosley78,rayet95}, the rp process \cite{Schatz98, Schatz01},
the $pn$ process \cite{Goriely}, and the $\nu p$ process 
\cite{Froehlich}. The $p$ nuclei $^{180}$Ta and $^{138}$La may 
additionally need contributions from a $\nu$ process \cite{Woosley90}. 
A recent study \cite{travaglio} found that light $p$ nuclei are 
produced in sufficient amounts to explain the abundances in the 
solar system by a combination of proton capture reactions and 
photodisintegrations in the thermonuclear explosion of a white 
dwarf (type Ia supernova), although earlier simulations did not 
predict such a production \cite{howmeywoo,howmey,kusa05}.

In total, these processes include several thousands of reactions on 
thousands of mainly unstable nuclei. This indicates already that it 
is not possible to measure all these reaction rates in the laboratory. 
Therefore, with very few exceptions, the astrophysical reaction rates 
are calculated by means of the statistical Hauser-Feshbach model 
\cite{HauserFeshbach}. Nuclear physics input, such as ground state (g.s.) properties, 
level densities, $§\gamma$-ray strength functions, and optical 
model potentials are needed for these calculations. This input can be 
determined, tested, and improved by laboratory measurements to finally 
obtain robust and reliable model predictions for these parameters.

It was pointed out \cite{Rapp, Rauscher} that certain reactions 
are of particular importance in the $\gamma$-process reaction network, 
because a change in their reaction rate has a direct impact on the 
calculated $p$ nuclei abundance. One of these particular important 
reactions is the proton capture reaction on $^{74}$Ge. It is the 
first reaction in the sequence 
$^{74}$Ge(p,$\gamma$)$^{75}$As(p,n)$^{75}$Se($\gamma$,n)$^{74}$Se, 
a main production route to the $p$ nucleus $^{74}$Se. A strong reaction 
flow is found to proceed through this sequence as illustrated in 
Fig.~\ref{fig:flow}, showing the results of our $\gamma$-process 
calculation in a 25 M$_\odot$ star using similar conditions as in 
Refs.~\cite{rayet95,Rapp} but employing updated reaction rates.

The reaction $^{74}$Ge(p,$\gamma$)$^{75}$As was investigated using 
the in-beam technique, where the $\gamma$ decay after proton capture 
is analyzed with high-resolution high-purity germanium detectors 
\cite{Harissopulos, Galanopoulos}. This method
makes it possible to investigate besides the total cross section the partial 
cross section for the direct population of several states in $^{75}$As.
This is an advantage compared to the widely used activation technique
\cite{Yalcin, Filipescu, Dillmann, Kiss, Sauerwein, Halasz}, which was
not applicable anyway since the reaction product $^{75}$As is stable. 

The method followed here was already demonstrated by Galanopoulos 
\textit{et al.} \cite{Galanopoulos} of being very powerful to determine 
very small cross sections, down to $1~\mu \mathrm{b}$ for astrophysics 
applications. This makes possible an investigation of the $^{74}$Ge(p,$\gamma$) reaction between 
$E_{p}=2.1 ~\mathrm{MeV}$ and $E_{p}= 3.7 ~\mathrm{MeV}$.

%Since the product of the $^{74}$Ge(p,$\gamma$)$^{75}$As reaction
%is stable, the widely used activation technique 
%\cite{Yalcin, Filipescu, Dillmann, Kiss, Sauerwein, Halasz}
%is not applicable. Therefore, we used the in-beam technique where 
%the $\gamma$ decay after proton capture is analyzed with high-resolution 
%high-purity germanium detectors \cite{Harissopulos, Galanopoulos}. 

\section{Astrophysically relevant energy window and sensitivities of $^{74}\mathrm{Ge}(\mathrm{p},\gamma)^{75}\mathrm{As}$}
\begin{figure}
 \includegraphics[width=\columnwidth]{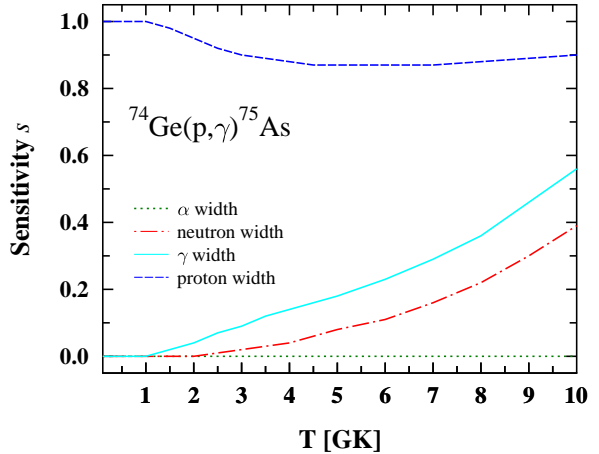}
 \caption{\label{fig:ratesensi} (Color online) Sensitivities $s$ of the
 astrophysical reaction rate for $^{74}$Ge(p,$\gamma$)$^{75}$As as a
 function of stellar plasma temperature $T$ when varying neutron, 
 proton, $\alpha$, and $\gamma$ widths separately by a factor of two. 
 The relevant temperature in the $\gamma$ process is close to
 $T=3~\mathrm{GK}.$}
 \end{figure}

It is very important to measure close to or inside the astrophysically 
relevant energy range to ensure that dependencies on nuclear properties 
are similar to those appearing in the stellar rates because these 
sensitivities strongly vary with energy. The sensitivities of the 
astrophysical reaction rate of $^{74}$Ge(p,$\gamma$)$^{75}$As to a 
change in the averaged $\gamma$, neutron, proton, and $\alpha$ widths 
used in the rate and cross-section models are shown in 
Fig.~\ref{fig:ratesensi}. Here, the averaged widths were varied 
separately by a factor of two. A sensitivity $s = 1$ indicates that 
the rate is changed by the same factor as the width, while $s = 0$ 
signifies that the width variation has no influence on the 
predicted rate. Details on the calculation of the sensitivity 
factor $s$ can be found in \cite{Rauscher_Sensitivity,sensi}.

As can be seen in Fig.~\ref{fig:ratesensi}, the rate prediction 
is almost exclusively sensitive to the proton width below about 
$4~\mathrm{GK}$, with the relevant stellar plasma temperature $T$ for 
this reaction in the $\gamma$ process being close to $3~\mathrm{GK}$. 
This rate sensitivity is related to the sensitivity of the reaction 
cross sections for $^{74}$Ge nuclei being in the g.s. 
and thermally excited states in the energy range mostly contributing 
to the reaction rate integral $\mathcal{I}$, appearing in the calculation 
of the stellar rate,

 \begin{equation}
 \label{eq:integral}
 \mathcal{I}=\int_0^\infty \sigma^\mathrm{eff}(E) \, \Phi(E,T) \, dE \quad,
 \end{equation}

where $\Phi(E,T)$ is the Maxwell-Boltzmann energy distribution of the 
projectiles and $\sigma^\mathrm{eff}(E)$ is the \textit{effective} cross 
section summing over all energetically possible transitions to states 
$\mu$, $\nu$ in the initial and final nucleus, respectively, 
\cite{Rauscher_Sensitivity}

 \begin{equation}
 \label{eq:effcs}
 \sigma^\mathrm{eff}=\sum_\mu \sum_\nu \frac{2J_\mu+1}{2J_0+1} \frac{E-E_\mu}{E}
 \sigma^{\mu \rightarrow \nu}(E-E_\mu) \quad.
 \end{equation}

Following Ref. \cite{fow74}, the above equation implicitly assumes that cross 
sections $\sigma^{\mu \rightarrow \nu}$ at zero or negative energies 
$E-E_\mu$ are zero. In general, the effective cross section includes 
more transitions than the laboratory cross section 
$\sigma^\mathrm{lab}=\sum_\nu \sigma^{0 \rightarrow \nu}$. However, 
it depends on the reaction and plasma temperature whether the additional 
transitions are contributing significantly. Their relative contribution 
to the stellar rate is given by $1-X$, where $X$ is the g.s.\ contribution 
to the stellar rate \cite{xfactor}

 \begin{equation}
 \label{eq:xfactor}
 X(T)=\frac{\int_0^\infty \sigma^\mathrm{lab}(E) \, \Phi(E,T) \, dE}{\int_0^\infty \sigma^\mathrm{eff}(E) \, \Phi(E,T) \, dE} \quad.
 \end{equation}

\begin{figure}
 \includegraphics[width=\columnwidth]{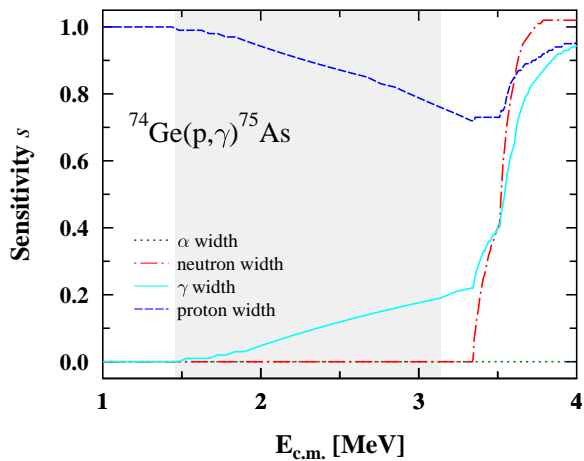}
 \caption{\label{fig:Sensitivity} (Color online) Sensitivities $s$ of the
 $^{74}$Ge(p,$\gamma$)$^{75}$As cross section when varying neutron, proton, 
 $\alpha$, and $\gamma$ widths separately by a factor of two. The
 astrophysical energy window for a temperature of $T = 3 ~\mathrm{GK}$ 
 is marked by the shaded area.}
 \end{figure}

For $^{74}$Ge(p,$\gamma$)$^{75}$As at $T=3$ GK, the integral $\mathcal{I}$ 
is dominated by contributions from energies $1.5\leq E\leq 3.1~\mathrm{MeV}$ \cite{Gamow}. 
Therefore, our measurement covers a significant fraction of the relevant 
energy range. However, the g.s.\ contribution to the stellar rate 
at $3~\mathrm{GK}$ is $X=0.66$; that is, only 2/3 of the stellar rate is obtained 
by integrating the laboratory cross section. Nevertheless, the sensitivity of 
$\sigma^\mathrm{lab}$ in the lower part of the measured energy range, shown 
in Fig.~\ref{fig:Sensitivity}, is similar to the one of the astrophysical 
reaction rate; that is, there is a dominant sensitivity to variations in the proton 
width. The calculated proton width depends on nuclear masses (determining the 
relative energies in all reaction channels), the level schemes, and the 
proton+nucleus optical potential \cite{Rauscher_Sensitivity,sensi}. Masses 
and level schemes for nuclei close to the valley of stability are known with 
sufficient accuracy. Hence, they are no major source of uncertainty in the 
calculation of the proton width. Therefore, this measurement of the 
$^{74}$Ge(p,$\gamma$) reaction is well suited to test the proton optical 
potential used in the prediction of the rate and to improve the astrophysical 
reaction rate at $\gamma$-process temperatures.

The experimental details of the measurement are presented in 
Sec.~\ref{sec:experiment}. After introducing the in-beam technique, the 
data analysis is presented in Sec.~\ref{sec:data_analysis}. The deduced total 
and partial cross sections are compared to theoretical predictions from 
Hauser-Feshbach statistical model calculations in Sec.~\ref{sec:results}, 
where also final conclusions regarding the astrophysical reaction rate 
are given.
%Rapp \textit{et al.} have investigated the reaction flux patterns of the $p$ process within the framework of a multimass zone Type II SN shock-front model for a 25
%$\mathrm{m}_\odot$ star.

\section{\label{sec:experiment}Experiment}

The reaction $^{74}$Ge(p,$\gamma$)$^{75}$As ($Q$ value: 
$(6898.94 \pm 0.95)~\mathrm{keV}$ \cite{NNDC}) was measured at seven energies
partly inside the Gamow window. The proton energies were varied between 
$2.1\leq E_{p} \leq 3.7~\mathrm{MeV}$.

\subsection{\label{sec:targetpreparation}Preparation and characterization of targets}
Germanium, with an enrichment in $^{74}$Ge of 97.8 \%, was evaporated in 
vacuum onto a $80~ \frac{\mathrm{mg}} {\mathrm{cm}^2}$-thick gold foil. 
This backing was thick enough to stop the proton beam completely to ensure 
a reliable charge collection. The germanium target was characterized by 
Rutherford back scattering (RBS). The areal density of the target is 
$(301 \pm 12)~ \frac{\mu \mathrm{g}}{\mathrm{cm}^2}$, which corresponds 
to an average energy loss of about $22~\mathrm{keV}$ and $14~\mathrm{keV}$ at 
$E_{p}= 2.1~\mathrm{MeV}$ and $E_{p}= 3.7~\mathrm{MeV}$, respectively. 
The energy losses were calculated using the SRIM code \cite{SRIM}. 
To study the homogeneity of the target thickness, the areal density was determined 
at many positions of the target area. All measurements yielded the same thickness 
within the statistical uncertainties. A second RBS measurement after the 
experiment excluded losses of target material and deterioration effects.

\subsection{Experimental setup}

 \begin{figure}
 \includegraphics[width=\columnwidth]{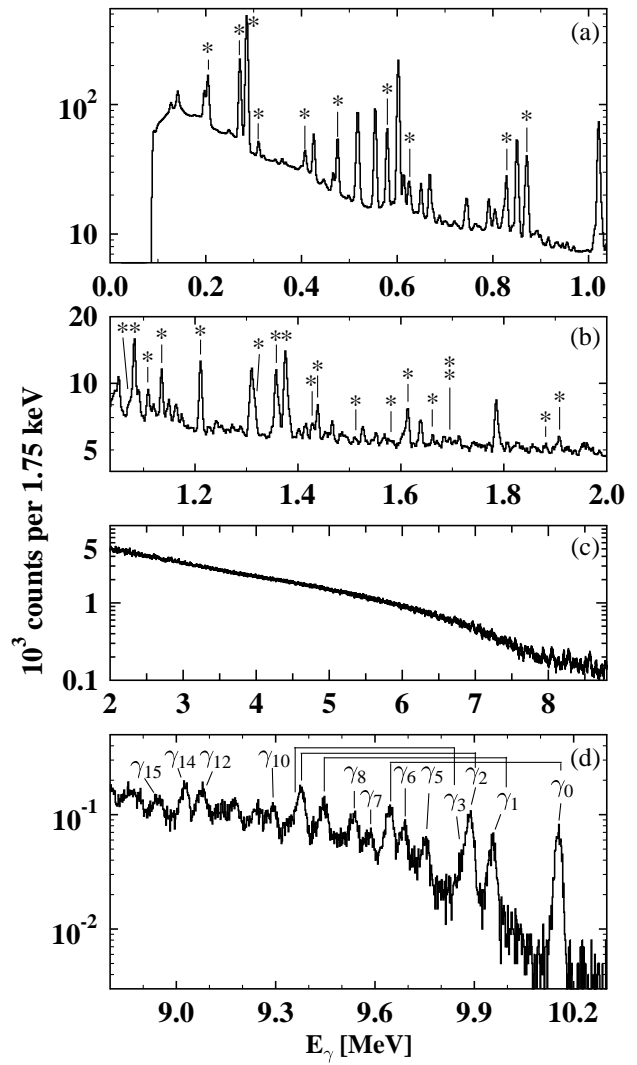}
 \caption{\label{fig:Spectrum} Typical spectrum while bombarding $^{74}$Ge
 with $3.3\textendash\mathrm{MeV}$ protons. This spectrum was recorded at $0^\circ$ relative
 to the beam axis. All transitions to the g.s. of the reaction product
 $^{75}$As are marked with an asterisk. Transitions from the so-called entry state
 to the g.s. or one of the excited states are marked with $\gamma_\mathrm{X}$. In 
 particular, the transition to the g.s. is denoted by $\gamma_0$, whereas
 the transition to the first excited state is denoted by $\gamma_1$ and so on.
 For some of these transitions the single escape peak is marked as well.}
 \end{figure}

The measurements were carried out at the 5.5 MV T11 Van de Graaff tandem accelerator 
of the Institute of Nuclear Physics of the National Center for Scientific Research 
(NCSR) "Demokritos". The accelerator was calibrated by means of the $992\textendash\mathrm{keV}$ 
resonance of the $^{27}$Al(p,$\gamma$)$^{28}$Si reaction. The air-cooled germanium 
target was bombarded for several hours with protons at beam currents of about 
$250~\mathrm{nA}$. The beam, which had a diameter of $4~\mathrm{mm}$, impinges 
perpendicular on the target. The beam current was determined by the charge deposited 
in the backing with a current integrator with an uncertainty of less than 4 \%. 
A negatively charged diaphragm $(U = - 400 \mathrm{V})$ suppresses secondary electrons at the 
entrance of the target chamber.

The proton capture reactions were identified by detecting the prompt $\gamma$ decays of 
the reaction products with four high-purity germanium (HPGe detectors). Three detectors have a relative 
detection efficiency of 100\%, whereas the fourth detector has a relative efficiency of 
80\%  compared to a $7.62~\times~7.62\textendash\mathrm{cm}$ cylindrical NaI detector 
at $E_\gamma= 1.33 ~ \mathrm{MeV}$. The detectors were mounted as close as
possible around the target chamber at a distance of $16~\mathrm{cm}$ on a turnable 
table under fixed angles of 0$^{\circ}$, 90$^{\circ}$, 190$^{\circ}$, and 305$^{\circ}$ 
relative to the beam axis. After each measurement, this table was rotated by an angle of 
15$^{\circ}$ and the target was bombarded with the same energy to measure at four 
further angles of 15$^{\circ}$, 105$^{\circ}$, 205$^{\circ}$, and 320$^{\circ}$ 
relative to the beam axis. The proton beam impinged also on a blank gold backing for each 
energy and each table position, to investigate possible yield contributions from
reactions occurring in the backing material. 
Singles spectra were taken using the same target in all measurements. A typical 
spectrum for a proton energy of $3.3~\mathrm{MeV}$ is shown in Fig.~\ref{fig:Spectrum}(a)-\ref{fig:Spectrum}(d). 
This spectrum was recorded with a detector at $0^\circ$ relative to the beam axis.
Because of the large distance between detector and target summing effects of the $\gamma$ rays 
are negligible ($<$ 1\%), which was checked with a GEANT4 simulation \cite{GEANT4}.

%This procedure was repeated at each energy point with the proton beam impinging on a blank gold backing, in order

\section{\label{sec:data_analysis}Data analysis}
 \subsection{Production of the compound nucleus}
 \begin{figure}
 \includegraphics[width=\columnwidth]{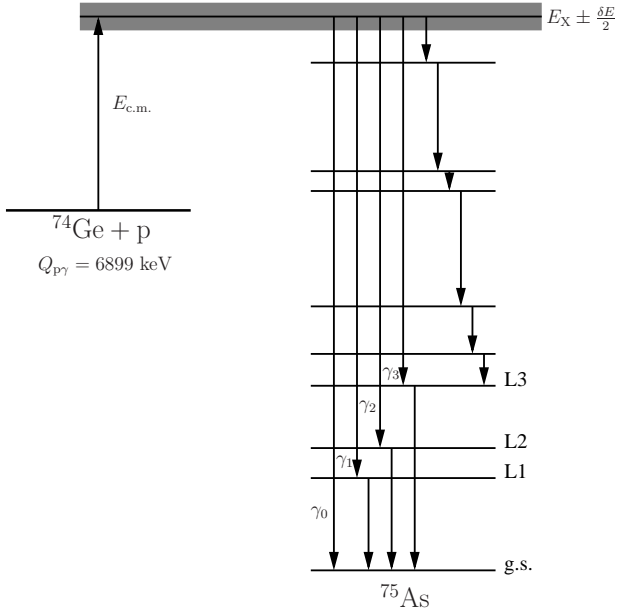}
 \caption{\label{fig:reaction} Schematic illustration of the reaction mechanism and decay. The target 
 nucleus is bombarded with protons to produce a compound nucleus in an excited state with energy $E_\mathrm{X}$. 
 $E_\mathrm{c.m.}$ denotes the center-of-mass energy of the system. An entry state is de-excited either via a direct $\gamma$ ray to 
 the ground or to excited states (denoted as $\gamma_0$, $\gamma_1$, \dots). Each observed $\gamma$ ray
 is actually a sum of the $\gamma$ rays emitted by all entry states around $E_\mathrm{X}$.}
 \end{figure}

To understand the data analysis of this in-beam experiment, it is useful to discuss the capture mechanism 
which leads to the formation of a compound nucleus. The target is bombarded with protons of
energy $E_{p}$ to form a highly excited compound nucleus at the energy 
$E_\mathrm{X} = S_{p} + E_\mathrm{c.m.}$, where $S_{p}$ is the proton separation energy in the 
compound nucleus and $E_\mathrm{c.m.}$ is the center-of-mass energy. The proton separation energy is equal 
to the reaction $Q$ value of proton capture. The reaction mechanism and the decays are illustrated in Fig.~\ref{fig:reaction}.
In practice, there is a large number of resonances which cannot be resolved within the energy uncertainty 
$\delta E$ when the nuclear level density is sufficiently high. This is fulfilled in $^{75}$As at 
basically all projectile energies. The energy uncertainty $\delta E$ depends on the spread of the initial
beam energy and the energy loss in the target. In this experiment the energy uncertainty is about $\pm 7~\mathrm{keV}$ for
all energies. Therefore, not only one discrete entry state is excited but 
a large number of states in a small energy interval. The measurement includes the formation and decay of 
all these entry states simultaneously.

The g.s. can be populated by a single $\gamma$-ray transition directly from the entry energy as 
well as by cascades of $\gamma$ rays populating and depopulating various excited levels, as sketched 
in Fig.~\ref{fig:reaction}. The former transition is the so-called $\gamma_0$ transition with an energy 
$E(\gamma_0) \approx E_\mathrm{X}$. The direct feeding of the first excited level $L_1$ or of the 
second excited level $L_2$ results in the $\gamma_1$ transition or in the $\gamma_2$ transition, respectively, 
and so on. For each entry state the total $\gamma$ width $\Gamma_\gamma$ is the sum of 
all partial widths $\Gamma_0$, $\Gamma_1$, \dots, to the final states, $\Gamma_\gamma=\Gamma_0+\Gamma_1+\dots$. 
The observed strengths and widths of the $\gamma$-ray transitions to discrete states, in turn,
are actually the sum of the discrete transitions from each entry state around $E_\mathrm{X}$ which cannot be resolved individually.

The statistical Hauser-Feshbach model of compound reactions -- used in the prediction of astrophysical reaction rates 
for this reaction -- uses the fact that the compound resonances cannot be resolved when encountering many states 
with widths larger than their spacing. Instead of considering the sums of hundreds of resonances with tiny energy 
differences, it is equivalent \cite{HauserFeshbach,Rauscher_Sensitivity} to calculate the formation and decay 
widths by assuming that states with all spins and parities are present at each $E_\mathrm{X}$, with widths representing 
the \textit{average} of the widths of all overlapping resonances within $\delta E$. These are the averaged widths used in 
the sensitivity study shown in Sec.~\ref{sec:introduction} and also used in the
theoretical discussion of the experimental results in Sec.~\ref{sec:results}.

\subsection{Determination of the total cross sections}
\label{total_WQS_inbeam}
 \begin{figure}
 \includegraphics[width=\columnwidth]{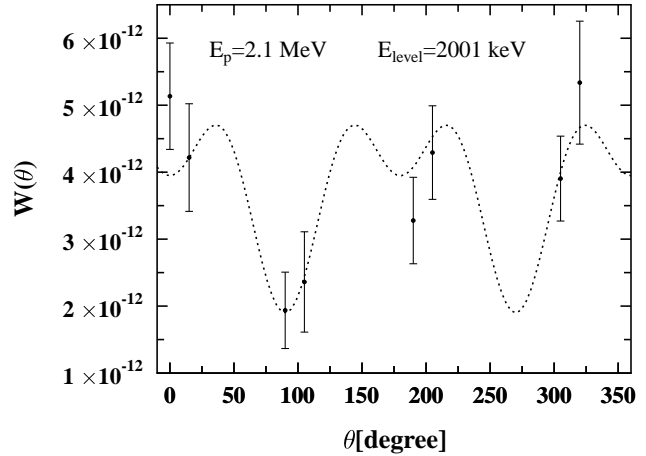}
 \caption{\label{fig:Winkelverteilung} Angular distribution for the
 $\gamma$-ray transition from the level with energy
 $E_\mathrm{level}=2001~\mathrm{keV}$ to the g.s. for an incident 
 energy of $E_p=2.1~\mathrm{MeV}$. The dotted curve  corresponds 
 to the Legendre polynomial fitted to the data points.}
 \end{figure}

To determine the total cross section of a capture reaction leading 
to the formation of a compound nucleus, the absolute number of produced compound 
nuclei $N_\mathrm{comp}$ has to be determined:

 \begin{equation}
 N_\mathrm{comp}= \sigma \cdot N_p \cdot m_T ~.
 \end{equation}

\begin{figure}
\includegraphics[width=\columnwidth]{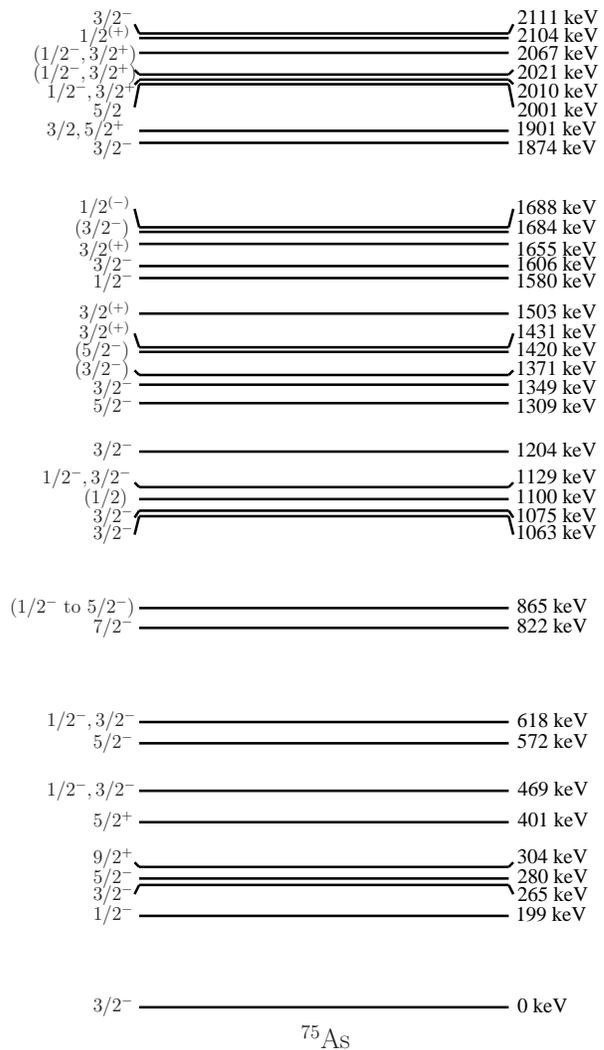}
\caption{\label{fig:levels} Partial level scheme of $^{75}$As showing all those levels from which transitions 
to the g.s. were observed in this experiment. All these transitions were taken into
account in the determination of the total cross section. The excitation energies $E_\mathrm{X}$, 
as well as the spins and parities shown were adopted from the compilation of Ref.~\cite{ENSDFnew}.}
\end{figure}

The absolute number of projectiles $N_p$ can be calculated from the accumulated charge and 
the areal particle density in $\frac{1}{\mathrm{cm}^2}$ is denoted as $m_T$. The absolute number of 
produced compound nuclei was obtained for each beam energy by measuring the angular distributions 
of all $\gamma$ rays to the g.s. of the produced compound nucleus. This is 
done by measuring the intensities $Y(E_\gamma)$ of all the relevant $\gamma$-ray transitions at eight angles $\theta$ relative to the beam axis. 
First, these intensities are normalized to the corresponding number of projectiles and the absolute 
full-energy efficiency of the respective detector $\epsilon(E_\gamma)$. The resulting normalized intensities are further 
corrected by the ratio between live and real time $\tau = \frac{t_\mathrm{LIVE}}{t_\mathrm{REAL}}$ of the 
data acquisition system. 
%For clarity, these normalized intensities are named $W(\theta)$. 
For a certain detector, which is mounted at an angle $\theta_A$ relative to the beam axis, and a g.s. transition with energy $E_\gamma$,
this normalized intensity is named 
\begin{equation}
 W(\theta_A)= \frac{Y(E_\gamma)}{\epsilon(E_\gamma) ~ \tau ~ N_p} \quad.
\end{equation}
Hence, at each beam energy an angular distribution for each $\gamma$-ray transition is obtained by fitting by a sum of Legendre polynomials to the eight experimental $W(\theta_X)$ values:
\begin{equation}
W(\theta) = A_0^i \left(1 + \sum_{k} \alpha_k P_k\left(cos\theta\right)\right)~~~(k=2,4,..) \quad.
\end{equation}
The coefficients $\alpha_k$, which are adjusted to reproduce the experimental angular distributions, 
are energy dependent. The maximum value of index $k$ depends on the multipolarity of the $\gamma$-ray 
transition in consideration. The scaling factor $A_0^i$ is proportional to the number of produced 
compound nuclei, which are deexcited via a cascade involving the g.s. transition $i$.
As an example, Fig.~\ref{fig:Winkelverteilung} shows the angular distribution
for the $\gamma$-ray transition from the $2001\textendash\mathrm{keV}$ level to the g.s. for an incident 
energy of $E_p=2.1~\mathrm{MeV}$. 
For every $\gamma$-ray transition $i$ to the g.s. of the produced 
compound nucleus, an $A_0^i$ coefficient is obtained. The total number of coefficients is $N$. The 
absolute number of produced compound nuclei per projectile is equal to the number of emitted $\gamma$ rays 
directly to the g.s. per projectile into the whole solid angle and can be derived from
\begin{equation}
 \label{NComp}
 \frac{N_\mathrm{comp}}{N_{p}} = \sum^{N}_{i=1} A_0^i \quad.
\end{equation}
The total cross section is then given by the expression:
\begin{equation}
\label{Sigma}
\sigma = \frac{\sum^{N}_{i=1} A_0^i}{m_T} ~.
\end{equation}
As can be seen in Eqs. (\ref{NComp}) and (\ref{Sigma}), only the scaling factor $A_0^i$ contributes to the 
total cross section. The shape of the angular distribution does not contribute. This is attributable to the fact that 
an integral over the whole solid angle of a sum of Legendre polynominals vanishes. \\
In this experiment, 35 g.s. transitions were observed for each angle. The corresponding levels are 
depicted in Fig.~\ref{fig:levels}. In total, 1960 peaks in the spectra had to be analyzed to obtain the 
total cross section of this reaction for seven energies.

\subsection{Determination of the partial cross sections}
It has to be emphasized that owing to the high efficiency of
the setup, several $\gamma$-ray transitions, which depopulate 
the entry state, were observed. For each of these $\gamma$-ray
transitions $j$ the angular distribution were determined with 
the procedure described in Sec.~\ref{total_WQS_inbeam}. From 
the coefficients $A_0^j$ the partial cross sections were 
determined by:

 \begin{equation}
 \sigma^j = \frac{A_0^j}{m_T} ~.
 \end{equation}

In total, 12 partial cross sections were deduced in this way.

\subsection{Determination of the detector efficiencies}
For the calculation of the total and partial cross sections the absolute full-energy efficiencies of 
the four detectors, at two angles each, have to be known. Because the efficiencies were needed up to 
energies of about $12~\mathrm{MeV}$, besides calibration sources the $^{27}$Al(p,$\gamma$) reaction was used 
for the efficiency calibration. For this purpose, the resonance at $E_{p}$ = 992 keV 
was used. For the $^{27}$Al(p,$\gamma$) reaction relative efficiency curves were 
first determined using the branchings reported in Ref.~\cite{27Al}. These branchings agree with those 
reported in Ref.~\cite{Endt} within 5 $\%$. The obtained relative efficiency curves from this reaction were 
matched to the absolute efficiencies gained with calibration sources.

\section{\label{sec:results} Results and Discussion}

% \subsection{Total cross sections}
% \begin{table}
% \caption{\label{tab:cross_section} Summary of the experimental $^{74}$Ge(p,$\gamma$) cross sections listed for each %center-of-mass energy $\mathrm{E}_\mathrm{cm}$.}
% \begin{ruledtabular}
% \begin{tabular}{cc}
%  \textrm{$\mathrm{E}_\mathrm{cm}$} & $\sigma$ \\
%  keV &  mb \\
% \colrule
%                     &              \\
% 2063 $\pm$ 7  & $0.65 \pm 0.06 $\\
%                     &              \\
% 2360 $\pm$ 7  & $1.93 \pm 0.16 $\\
%                     &              \\
% 2657 $\pm$ 7  & $4.50 \pm 0.38 $\\
%                     &              \\
% 2955 $\pm$ 7  & $8.27 \pm 0.70 $\\
%                     &              \\
% 3251 $\pm$ 7  & $14.06 \pm 1.20$\\
%                     &              \\
% 3449 $\pm$ 7  & $13.85 \pm 1.19$\\
%                     &              \\
% 3646 $\pm$ 8  & $5.69 \pm 0.58 $\\
% \end{tabular}
% \end{ruledtabular}
% \end{table}

\subsection{Total cross sections}

\subsubsection{Experimental results}

\begin{table*}
\caption{\label{tab:cross_section} Summary of the experimental $^{74}$Ge(p,$\gamma$) cross sections listed for each center-of-mass energy $E_\mathrm{c.m.}$.}
\begin{ruledtabular}
\begin{tabular}{c|c c c c c c c}
 \textrm{$E_\mathrm{c.m.}$} [keV] & 2063 $\pm$ 7 & 2360 $\pm$ 7  & 2657 $\pm$ 7  & 2955 $\pm$ 7 & 3251 $\pm$ 7 & 3449 $\pm$ 7 & 3646 $\pm$ 8 \\
 $\sigma$ [mb] &  $0.65 \pm 0.06 $ &  $1.93 \pm 0.16 $ & $4.5 \pm 0.4 $ & $8.3 \pm 0.7 $ & $14.1 \pm 1.2$ & $13.9 \pm 1.2$ &  $5.7 \pm 0.6 $\\
\end{tabular}
\end{ruledtabular}
\end{table*}

\begin{figure}
\includegraphics[width=\columnwidth]{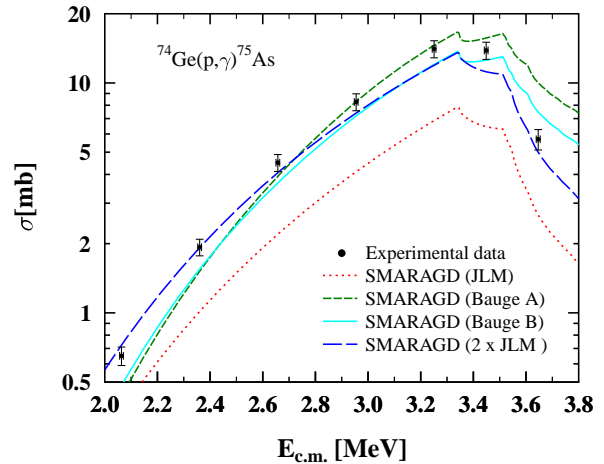}
\caption{\label{fig:WQS} (Color online) Total cross sections for the reaction $^{74}$Ge(p,$\gamma$)$^{75}$As 
as a function of center-of-mass energy. The total experimental cross sections are compared to predictions with the SMARAGD 
code \cite{SMARAGD} using different averaged proton widths from \cite{JLM} with low-energy modifications of 
\cite{Lejeune} (JLM), \cite{bauge} (Bauge A), and \cite{baugelane} (Bauge B). Furthermore, the 
experimental values are compared to predictions using the doubled value of the proton widths from (JLM), 
denoted as (2 $\times$ JLM) in the figure. }
\end{figure}

The experimental total cross sections are given in Table~\ref{tab:cross_section} and shown in Fig.~\ref{fig:WQS}.
The experimental values are not corrected for electron screening \cite{Claudrons,Rolfs}. 
An electron screening correction, similar to the one used in Ref.~\cite{Kettner,Spyrou}, would 
lead to an decreasing of the cross section of about 5~\% and 2~\% for the lowest and highest energies measured
in this experiment, respectively. 
The energy $E_{p}$ of the protons was obtained by correcting the adjusted primary energy $E_0 $ 
of the proton beam with the average energy loss $\Delta E$ inside the target material
\begin{equation}
 E_{p} = \frac{E_0+\left( E_0-\Delta E\right) }{2} \quad.
\end{equation}
This energy determination, which is appropriate if the energy loss and the cross section change only slightly 
over the target thickness, is independent of a specific cross section prediction. In this case, the uncertainties 
of the cross sections are larger than the changes in the cross section over the target thickness. The widths of 
the energies are determined by the uncertainty of the beam energy of the accelerator and the straggling in the 
target. They were added according to Gaussian error propagation. The energy uncertainty of the tandem accelerator 
at Demokritos was determined by scanning the resonance in $^{27}$Al(p,$\gamma$)$^{28}$Si  at 992 keV. An energy 
uncertainty of $\pm 3.8 ~ \mathrm{keV}$ was found. The energy loss of the protons in the target itself is obtained 
by a GEANT4 \cite{GEANT4} simulation and results in $(14 - 22) ~ \mathrm{keV}$, which agrees with the
values obtained with SRIM (compare Sec.~\ref{sec:targetpreparation}). The distribution of the proton beam 
after traveling through the target material is described by this simulation. The maximum of this distribution
is $E_0-\Delta E$ and the half width of this distribution at $1/e$ of the maximum is the energy 
straggling in the target which was about $6~\mathrm{keV}$ for all energies. Note that 
Table~\ref{tab:cross_section} gives center-of-mass energies to facilitate comparison with theoretical calculations.

\subsubsection{Implications for astrophysical proton capture on $^{74}$Ge}
\label{sec:theorytot}

In Fig.~\ref{fig:WQS}, the experimental values are compared 
to statistical model calculations performed with 
the code SMARAGD \cite{SMARAGD}, which includes more current level schemes \cite{ENSDFnew} and an improved calculation 
method compared to the NON-SMOKER code used in Ref.~\cite{ADNDT}. As discussed in Sec.~\ref{sec:introduction} and shown 
in Fig.~\ref{fig:Sensitivity}, the capture cross sections below the (p,n) threshold are mainly sensitive to the proton 
width and the only uncertainty in this width stems from the optical proton+nucleus potential applied in the calculation. The 
potential from Ref.~\cite{JLM} with low-energy modifications introduced in Ref.~\cite{Lejeune} is used as a standard potential.
It is named `JLM' in Fig.~\ref{fig:WQS}. The JLM potential yields 
an underestimation of the measured data by almost a factor of two, while the overall energy dependence is 
described well, also at the opening of the (p,n) channel where width fluctuation corrections have to be applied.

Exploratory calculations with different optical potentials and input to the potentials were performed. The microscopic JLM 
potential requires a nuclear matter density as input. The standard calculation shown here 
makes use of the nuclear matter density HFB-02 \cite{hfb02}. Using density distributions from the droplet model from HFB-14 \cite{hfb14} 
or from Ref.~\cite{negele} slightly changed the energy dependence but did not affect the magnitude of the cross section 
significantly. Recently, a modification of the imaginary part of the JLM potential was suggested to 
explain the energy dependence of low-energy (p,$\gamma$) data \cite{Rauscher_Sensitivity,ungarn1,ungarn2,ungarn3}. 
Again, the impact here is negligible and it cannot account for the discrepancy of a factor of two. A comparison of 
the $^{74}$Ge(p,$\gamma$)$^{75}$As data to results obtained with two additional optical potentials is shown in 
Fig.~\ref{fig:WQS} as well. Both are a reparametrization of the potential by Ref.~\cite{JLM} using modern scattering 
data, with one (case A, \cite{bauge}) being directly based on Ref.~\cite{JLM} and the other (case B, \cite{baugelane}) 
being a Lane-consistant version. These potentials are named 'Bauge A' and 'Bauge B' in Fig.~\ref{fig:WQS}. Unfortunately, both do not include low-energy modifications as suggested by 
Ref.~\cite{Lejeune}. As seen in Fig.~\ref{fig:WQS}, both reparametrizations give a better estimation of the absolute 
magnitude of the cross sections in the measured energy region but they exhibit a different energy dependence. 
A good reproduction of the energy dependence, however, is crucial for the prediction of the astrophysical 
reaction rate, as relevant contributions to the reaction rate integral defined in Eq.~\eqref{eq:integral} also 
come from energies lower than the measured ones.

Because the standard JLM potential accounts for the correct 
energy dependence of the cross sections, in the following 
we use the proton widths of this potential but multiply 
them by a factor of two. The resulting cross sections 
are also shown in Fig.~\ref{fig:WQS}. We obtain ratios
between the SMARAGD predictions using the increased proton 
widths and the experimental (p,$\gamma$) cross sections close 
to unity. At the proton energies for which the cross sections 
are mainly sensitive to the proton widths they are $0.89 \pm 0.08$, 
$1.00 \pm 0.08$, and $1.10 \pm 0.09$ at c.m. energies of
2.063 MeV, 2.360 MeV, and 2.657 MeV, respectively. As further discussed in the 
following section, this procedure is confirmed by the fact that 
also the partial cross sections are reproduced correctly.

\begin{figure}
\includegraphics[width=\columnwidth]{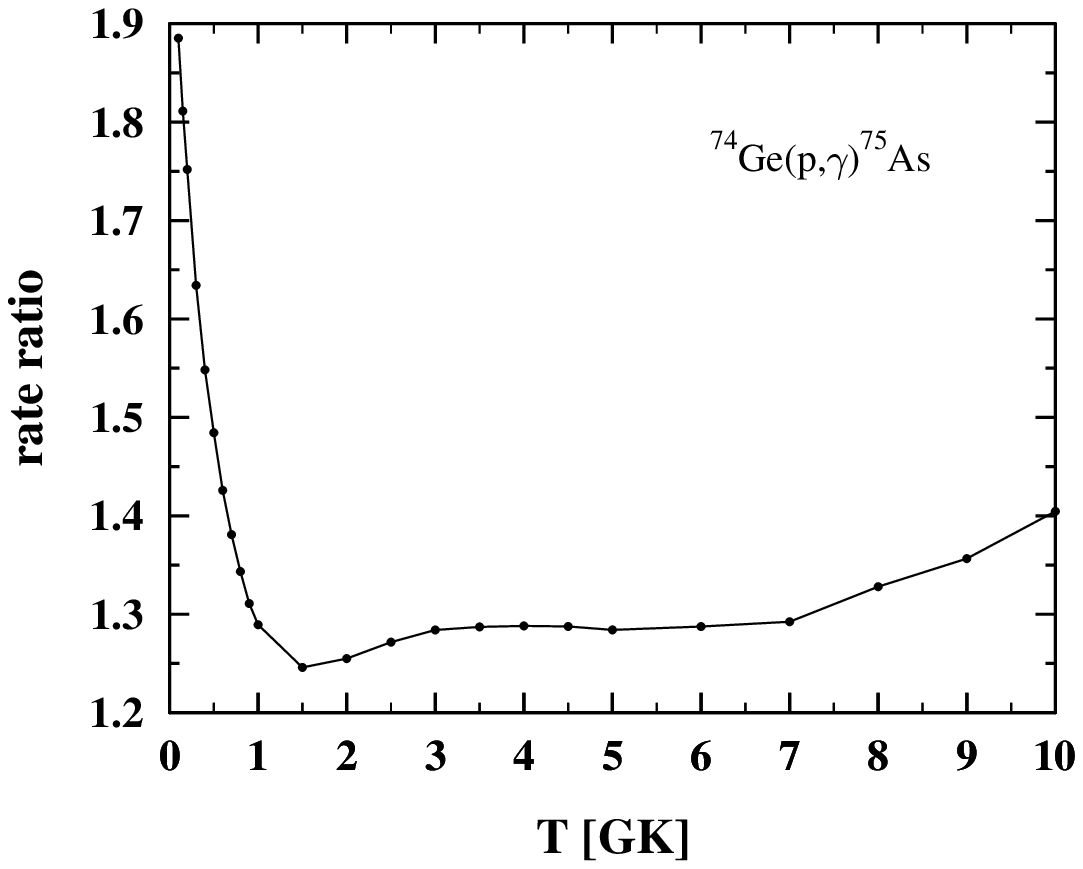}
\caption{\label{fig:pgratio} Ratio of the stellar rate from Table~\ref{tab:pgrate} and the rate given in Ref.~\cite{ADNDT}.}
\end{figure}

Assuming that the modification of the proton widths 
also applies to the additional transitions appearing in 
the calculation of the astrophysical reaction rate 
[Eqs.\ (\ref{eq:integral}) and (\ref{eq:effcs})], we 
show the new prediction of the proton capture reactivities 
in Table~\ref{tab:pgrate}. Those reactivitites are stellar 
reactivities in contrast to 
laboratory reactivities. 
The newly derived 
coefficients for the REACLIB parametrization \cite{adndt0} are given in Table~\ref{tab:pgcoeffs}. 
The ratio of the rate calculated with the new parameters to the one obtained from the tables 
of \cite{ADNDT} at different plasma temperatures is shown in Fig.~\ref{fig:pgratio}. At the 
relevant temperature of 
3 GK, the new rate is higher by 28\%. This is well within the expected uncertainty of the previous rate 
prediction and confirms earlier $\gamma$-process calculations \cite{rhhw02,hegXX,dillp} and 
their $^{74}$Se abundance.

\begin{table}
\caption{\label{tab:pgrate} Stellar reactivity $N_A \left< \sigma v \right>^*$ and g.s.\ contribution $X$ 
(taken from \cite{sensi}) for $^{74}$Ge(p,$\gamma$)$^{75}$As as a function of plasma temperature.}
\begin{ruledtabular}
\begin{tabular}{c c c c r@{$\times$}l c c c c}
$T$ & & & &\multicolumn{2}{c}{Reactivity} & & & &\textit{X} \\
$[\mathrm{GK}]$& & & & \multicolumn{2}{c}{[cm$^3$s$^{-1}$mole$^{-1}$]} & & & &\\
\hline
 &  & & \multicolumn{2}{c}{$~$}& & &\\
 0.10 & $~$ &$~$ &$~$& 2.564 $~~$ & $10^{-22}$&$~$ & $~$ & $~$ & 1.00 \\
 0.15 & $~$ &$~$ &$~$& 2.264 $~~$ & $10^{-17}$&$~$ & $~$ & $~$ & 1.00 \\
 0.20 & $~$ &$~$ &$~$& 2.733 $~~$ & $10^{-14}$&$~$ & $~$ & $~$ & 1.00 \\
 0.30 & $~$ &$~$ &$~$& 1.912 $~~$ & $10^{-10}$&$~$ & $~$ & $~$ & 1.00 \\
 0.40 & $~$ &$~$ &$~$& 4.908 $~~$ & $10^{-8}$ &$~$ & $~$ & $~$ & 1.00 \\
 0.50 & $~$ &$~$ &$~$& 2.479 $~~$ & $10^{-6}$ &$~$ & $~$ & $~$ & 1.00 \\
 0.60 & $~$ &$~$ &$~$& 4.848 $~~$ & $10^{-5}$ &$~$ & $~$ & $~$ & 1.00 \\
 0.70 & $~$ &$~$ &$~$& 5.137 $~~$ & $10^{-4}$ &$~$ & $~$ & $~$ & 1.00 \\
 0.80 & $~$ &$~$ &$~$& 3.560 $~~$ & $10^{-3}$ &$~$ & $~$ & $~$ & 1.00 \\
 0.90 & $~$ &$~$ &$~$& 1.809 $~~$ & $10^{-2}$ &$~$ & $~$ & $~$ & 1.00 \\
 1.00 & $~$ &$~$ &$~$& 7.272 $~~$ & $10^{-2}$ &$~$ & $~$ & $~$ & 1.00 \\
 1.50 & $~$ &$~$ &$~$& 8.946 $~~$ & $10^{0}$  &$~$ & $~$ & $~$ & 0.95  \\
 2.00 & $~$ &$~$ &$~$& 1.644 $~~$ & $10^{2}$  &$~$ & $~$ & $~$ & 0.86 \\
 2.50 & $~$ &$~$ &$~$& 1.175 $~~$ & $10^{3}$  &$~$ & $~$ & $~$ & 0.75 \\
 3.00 & $~$ &$~$ &$~$& 4.738 $~~$ & $10^{3}$  &$~$ & $~$ & $~$ & 0.66 \\
 3.50 & $~$ &$~$ &$~$& 1.300 $~~$ & $10^{4}$  &$~$ & $~$ & $~$ & 0.60 \\
 4.00 & $~$ &$~$ &$~$& 2.718 $~~$ & $10^{4}$  &$~$ & $~$ & $~$ & 0.55 \\
 4.50 & $~$ &$~$ &$~$& 4.674 $~~$ & $10^{4}$  &$~$ & $~$ & $~$ & 0.51 \\
 5.00 & $~$ &$~$ &$~$& 6.947 $~~$ & $10^{4}$  &$~$ & $~$ & $~$ & 0.49 \\
 6.00 & $~$ &$~$ &$~$& 1.124 $~~$ & $10^{5}$  &$~$ & $~$ & $~$ & 0.44 \\
 7.00 & $~$ &$~$ &$~$& 1.357 $~~$ & $10^{5}$  &$~$ & $~$ & $~$ & 0.40 \\
 8.00 & $~$ &$~$ &$~$& 1.328 $~~$ & $10^{5}$  &$~$ & $~$ & $~$ & 0.35 \\
 9.00 & $~$ &$~$ &$~$& 1.111 $~~$ & $10^{5}$  &$~$ & $~$ & $~$ & 0.30 \\
10.00 & $~$ &$~$ &$~$& 8.357 $~~$ & $10^{4}$  &$~$ & $~$ & $~$ & 0.24
\end{tabular}
\end{ruledtabular}
\end{table}

\begin{table}
\caption{\label{tab:pgcoeffs}REACLIB parameters for $^{74}$Ge(p,$\gamma$) and its reverse reaction, 
obtained from fitting the reactivities shown in Table~\ref{tab:pgrate}.}
\begin{ruledtabular}
\begin{tabular}{ccc}
Parameter & \multicolumn{1}{c}{(p,$\gamma$)} & \multicolumn{1}{c}{($\gamma$,p)} \\
\hline
       &                         &                        \\
%$a_0$ & -2.380584\times 10^{2} & 2. & 3. & 4. & 5. & 6.
$a_0$  & $2.780078\times 10^{1}$ & $5.010011\times 10^{1}$ \\
$a_1$  &  0.000000 & $-8.005979\times 10^{1}$ \\
$a_2$  & \multicolumn{2}{c}{$-5.063646\times 10^{1}$}  \\
$a_3$  &  \multicolumn{2}{c}{$2.235434\times 10^{1}$}   \\
$a_4$  & \multicolumn{2}{c}{$-2.185726$}  \\
$a_5$  & \multicolumn{2}{c}{$4.255545\times 10^{-2}$}  \\
$a_6$  & $-9.282484$ & $-7.782484$
\end{tabular}
\end{ruledtabular}
\end{table}

\subsection{Partial cross sections}
\label{sec:partial}

\subsubsection{Experimental results}

\begin{table*}
\caption{\label{tab:partial_cross_section} Summary of the experimental partial cross sections listed for each center-of-mass energy $E_\mathrm{c.m.}$. For comparison partial 
cross sections from Ratkevich \textit{et al.} \cite{Ratkevich} are given as well. Please note the large uncertainties in the energy. This data are marked with asteriks.}
\begin{ruledtabular}
\begin{tabular}{c|c@{$\pm$}l c@{$\pm$}l  c@{$\pm$}l c@{$\pm$}l c@{$\pm$}l c@{$\pm$}l}
 &\multicolumn{10}{c}{Energy of level which is directly fed from the ``entry state``}\\
 $E_\mathrm{c.m.}$&\multicolumn{2}{c}{0 keV} & \multicolumn{2}{c}{199 keV} & \multicolumn{2}{c}{265 keV} & \multicolumn{2}{c}{280 keV} & \multicolumn{2}{c}{401 keV} & \multicolumn{2}{c}{469 keV}\\
$[\mathrm{keV}]$ & \multicolumn{2}{c}{$\sigma(\gamma_0)$ [$\mu$b]} & \multicolumn{2}{c}{$\sigma(\gamma_1)$ [$\mu$b]} & \multicolumn{2}{c}{$\sigma(\gamma_2)$ [$\mu$b]} & \multicolumn{2}{c}{$\sigma(\gamma_3)$ [$\mu$b]} & \multicolumn{2}{c}{$\sigma(\gamma_5)$ [$\mu$b]} & \multicolumn{2}{c}{$\sigma(\gamma_6)$ [$\mu$b]}\\   \hline
 & \multicolumn{2}{c}{$~$} &\multicolumn{2}{c}{$~$} &\multicolumn{2}{c}{$~$} & \multicolumn{2}{c}{$~$}\\
2063 $\pm$ 7 	   &	11.8 $~$&0.9	 & 	9.8  $~$ &0.7  & 	9.6  $~$  &1.2	 & 	3.2  $~$ &1.0	 &\multicolumn{2}{c}{$~$}& 	7.9   $~$ &0.6 \\
2360 $\pm$ 7       &	20.8 $~$&1.5	 & 	19.0 $~$ &1.4  & 	30.6 $~$  &2.5	 & 	3.8  $~$ &1.0	 & 	7.1  $~$ &0.7	& 	14.7 $~$ &1.2 \\
*2400	$\pm$ 300  &	*22.4$~$&4.8	 & 	*19.8 $~$ &4.0   & 	*19.0$~$   &5.0	 &	 *3.7  $~$ &0.8	 & 	*7.8  $~$ &1.4	& 	*15.7  $~$ &3.2	\\
2657 $\pm$ 7       &	41.8 $~$&3.1	 & 	33.8 $~$ &2.5  & 	49.5 $~$  &3.9	 & 	12.0 $~$ &1.3	 & 	15.1 $~$ &1.4	& 	30.3  $~$ &2.4 \\
2955 $\pm$ 7 	   &	64.9 $~$&4.7	 & 	56.8 $~$ &4.1  & 	73.3 $~$  &5.6	 & 	19.8 $~$ &2.2	 & 	26.1 $~$ &2.1	& 	44.9  $~$ &3.4 \\
3251 $\pm$ 7 	   &	90.0 $~$&6.6	 & 	62.8 $~$ &4.8  & 	102.9$~$  &7.7	 & 	25.4 $~$ &2.7	 & 	39.6 $~$ &3.7	& 	64.9  $~$ &5.2 \\
3449 $\pm$ 7 	   &	82.6 $~$&6.1	 & 	58.6 $~$ &4.4  & 	90.5 $~$  &7.0	 & 	26.2 $~$ &3.0	 & 	30.6 $~$ &2.8	& 	45.1  $~$ &3.8 \\
3646 $\pm$ 8 	   &	31.0 $~$&2.3	 & 	21.7 $~$ &1.9  & 	32.2 $~$  &3.2	 & 	7.6  $~$ &1.0	 & 	10.3 $~$ &1.1	& 	19.9  $~$ &1.7 \\ \hline
 & \multicolumn{2}{c}{$~$} & \multicolumn{2}{c}{$~$} & \multicolumn{2}{c}{$~$} &  \multicolumn{2}{c}{$~$}\\
 &\multicolumn{10}{c}{Energy of level which is directly fed from the ``entry state``}\\
 $E_\mathrm{c.m.}$& \multicolumn{2}{c}{572 keV} & \multicolumn{2}{c}{618 keV} & \multicolumn{2}{c}{865 keV} & \multicolumn{2}{c}{1075 keV} & \multicolumn{2}{c}{1129 keV} & \multicolumn{2}{c}{1204 keV}\\
$[\mathrm{keV}]$ & \multicolumn{2}{c}{$\sigma(\gamma_7)$ [$\mu$b]} & \multicolumn{2}{c}{$\sigma(\gamma_8)$ [$\mu$b]} & \multicolumn{2}{c}{$\sigma(\gamma_{10})$ [$\mu$b]} & \multicolumn{2}{c}{$\sigma(\gamma_{12})$ [$\mu$b]} & \multicolumn{2}{c}{$\sigma(\gamma_{14})$ [$\mu$b]} & \multicolumn{2}{c}{$\sigma(\gamma_{15})$ [$\mu$b]}\\   \hline
 &\multicolumn{2}{c}{$~$} &\multicolumn{2}{c}{$~$} & \multicolumn{2}{c}{$~$}& \multicolumn{2}{c}{$~$}\\
2063 $\pm$ 7 	   &	2.0 $~$&0.3	 & 	5.7 $~$ &0.5 & 	3.1 $~$ &0.5	 	&  	8.0  $~$ &0.7	 & 	10.7 $~$ &0.9	 &	5.4  $~$ &0.5	\\
2360 $\pm$ 7       & 	3.2 $~$&0.5	 & 	12.4$~$ &1.2 & 	9.1 $~$ &0.9		& 	19.7 $~$ &1.6	 & 	26.5 $~$ &2.2	 &	13.6 $~$ &1.3	\\
*2400	$\pm$ 300  & 	*2.7 $~$&1.0	 &      *12.8$~$ &2.6  & 	*10.8$~$ &3.5	 	& 	*14.1 $~$ &3.2	 & 	*17.5 $~$ &1.4	 &	*10.2 $~$ &2.1	\\
2657 $\pm$ 7       & 	12.2$~$&1.4	 & 	25.7$~$ &2.3 & 	23.9$~$ &2.2	 	& 	37.2 $~$ &3.1	 & 	39.4 $~$ &3.2	 &	22.2 $~$ &2.3	\\
2955 $\pm$ 7       & 	16.4$~$&1.7	 & 	45.0$~$ &3.5 & 	34.5$~$ &2.9	 	& 	57.4 $~$ &4.4	 & 	63.1 $~$ &4.8	 &	44.0 $~$ &3.9	\\
3251 $\pm$ 7       & 	35.3$~$&3.4	 & 	60.9$~$ &5.9 & 	54.1$~$ &4.7	 	& 	78.2 $~$ &6.1	 & 	82.7 $~$ &6.6	 &	51.1 $~$ &4.5	\\
3449 $\pm$ 7       & 	17.9$~$&2.2	 & 	45.3$~$ &3.8 & 	43.9$~$ &3.9	 	& 	75.0 $~$ &6.0	 & 	87.0 $~$ &6.9	 &	41.8 $~$ &3.9	\\
3646 $\pm$ 8       & 	9.1 $~$&1.4	 & 	22.4$~$ &2.0 & 	27.4$~$ &2.5		& 	31.6 $~$ &2.9	 & 	27.0 $~$ &2.6	 &	20.6 $~$ &2.1	\\
\end{tabular}
\end{ruledtabular}
\end{table*}

\begin{figure*}
\includegraphics[width=\textwidth]{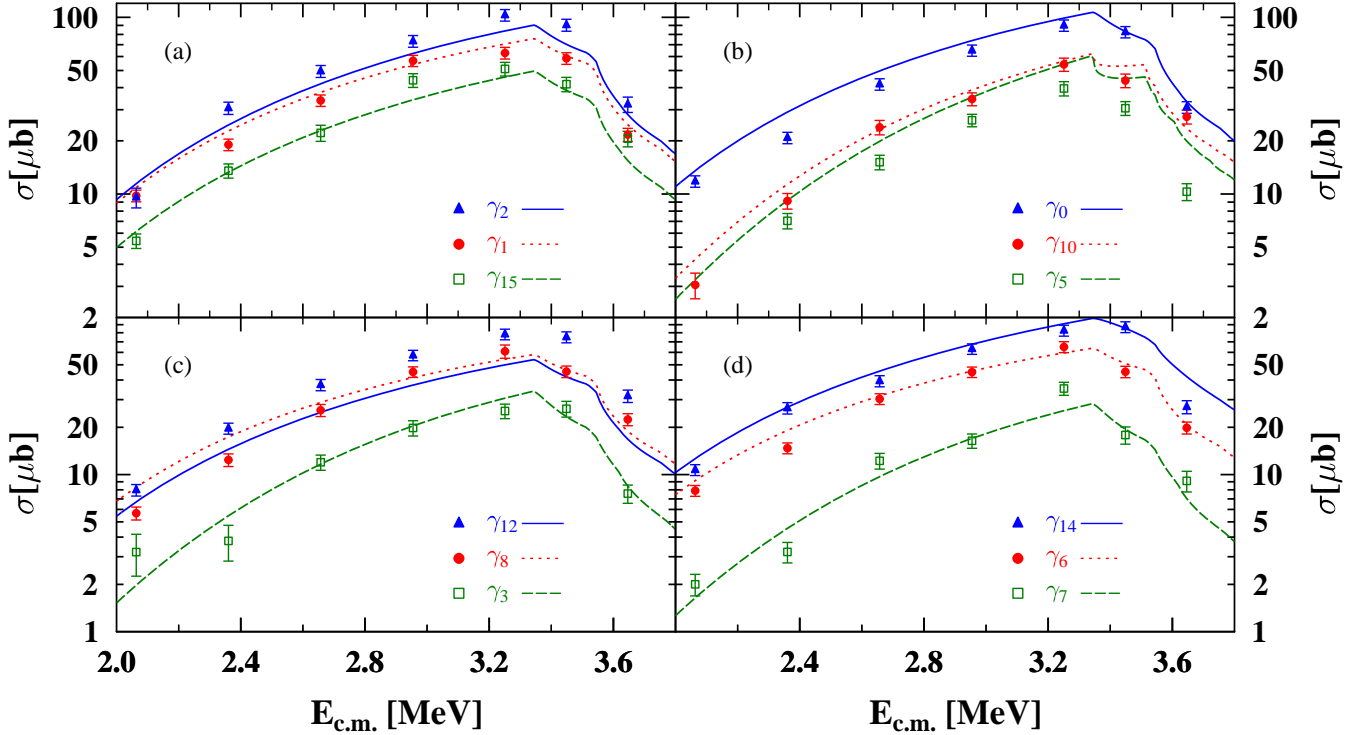}
\caption{\label{fig:PartialWQS} (Color online) Partial cross sections for the reaction $^{74}$Ge(p,$\gamma$)$^{75}$As as a
function of center-of-mass energy. Because the symbol widths are larger than the horizontal uncertainties, no horizontal error bars are depicted.
The experimental values are compared with predictions calculated with the SMARAGD code using the proton optical model potential 
of Refs.~\cite{JLM,Lejeune} multiplied by a factor of two.}
% The measured cross sections (symbols) are compared to the cross sections calculated using the standard proton+nucleus optical potential with the resulting proton 
%widths increased by a factor of two.}
\end{figure*}

In Table~\ref{tab:partial_cross_section} the measured partial cross sections are shown. Twelve transitions, 
which feed levels of $^{75}$As directly from the entry state, were observed and the partial cross 
sections for these levels were determined. The transition to the fourth excited state is strongly suppressed 
because of its spin and parity. The other three missing transitions to the $9^\mathrm{th}$, $11^\mathrm{th}$, 
and $13^\mathrm{th}$ state are visible in the $\gamma$ spectra, but the low peak-to-background ratio at the 
respective energies hamper a reliable analysis. Furthermore, a lot of $\gamma$-ray transitions to levels with
high excitation energy in $^{75}$As occur. Because the level scheme of $^{75}$As is incomplete in this energy range,
those transitions could not be analyzed.

Our results can be compared to values from Ref.~\cite{Ratkevich}. This reference 
gives a list of partial cross sections for a proton energy of $(2.4 \pm 0.3)~\mathrm{MeV}$. Please note the large uncertainties
for the energy. Their values are included in Table~\ref{tab:partial_cross_section}. In Ref.~\cite{Ratkevich} the primary $\gamma$-ray transitions were measured 
with a pair spectrometer consisting of a Ge(Li) detector which was surrounded by a four-section circular NaI(Tl) 
detector. The partial cross sections agree for most transitions with our values within the uncertainties. 
Nevertheless, there are slightly larger discrepancies for the partial cross sections populating the levels 
at 265, 1075, and 1129 keV.

\subsubsection{Comparison to theory and new spin assignments to the 401\textendash and 1075\textendash keV states in $^{75}$As}
\label{sec:partialspin}

The modification of the proton widths obtained with the JLM 
optical potential discussed in 
Sec.~\ref{sec:theorytot} also has to apply to the partial 
widths. As shown in Fig.~\ref{fig:PartialWQS}, the resulting 
cross section calculation (with the proton widths multiplied 
by a factor of two) agrees well with the data. There are 
two exceptions, discrepancies occur for the $\gamma_5$ and 
$\gamma_{12}$. These cannot be attributable to a problem in the 
optical potential as this would systematically affect also 
the other partial cross sections. Because all other partial 
cross sections are well reproduced, they have to be 
explained by incorrect spin or parity assignments.

The prediction of the partial cross sections for the 
$\gamma_{12}$ transition from the state at 1075 keV are 
too low. The spin assignment to this state is 3/2$^-$ 
\cite{ENSDFnew}. We find improved agreement when additionally 
including a 5/2$^-$ state at the same energy, 
leading to a doublet assignment of "3/2$^-$, 5/2$^-$".

The $\gamma_5$ case of the partial cross section of the state at 
401 keV is different because it is overpredicted by theory. This can 
only be remedied by changing the parity assignment from 5/2$^+$ to 5/2$^-$. 
The original 5/2$^+$ assignment quoted in Ref.~\cite{ENSDFnew} was 
obtained from $\ell=2$ partial waves in DWBA calculations 
\cite{betts74,schrader76,rotbard83}. It is conceivable that the 
data of \cite{betts74,schrader76,rotbard83} can also be reproduced
with $\ell=3$ but a further DWBA calculation is required for 
confirmation. It may be noted in passing that the original 
parity assignment was tentative \cite{nuccharts} but was adopted 
as proven in the subsequent papers, including the 
Refs.\ \cite{betts74,schrader76,rotbard83} cited by Ref.~\cite{ENSDFnew}.

%\begin{figure}
%\includegraphics[width=\columnwidth]{Partial_WQS_1128keV.eps}
%\caption{\label{fig:gamma14} (Color online)  Experimental partial cross sections feeding the state at 1128 keV are compared 
%to calculations assuming a spin and parity assignment of 1/2$^-$ and 3/2$^-$. Furthermore, a calculation assuming two 
%unresolved states with spin and parity of 1/2$^-$ and 3/2$^-$ at the same energy is shown.}
%\end{figure}

\subsubsection{Implications for astrophysical neutron capture on $^{74}$As}

The sensitivities of the partial cross sections to variations of the different widths are very similar to the ones shown 
in Fig.~\ref{fig:Sensitivity} for the total cross sections. There is a subtle difference, however, which can be used 
to extract additional information relevant for astrophysics.

The averaged widths $\left<\Gamma_x\right>$ of the possible reaction channels $x=\gamma,\mathrm{n},\mathrm{p},\alpha$ 
(i.e., in the $\gamma$, neutron, proton, and $\alpha$ channel) enter the calculation of the total (p,$\gamma$) cross section 
in the form of a sum over fractions \cite{Rauscher_Sensitivity,sensi}
\begin{equation}
\label{eq:totfrac}
\sigma \propto \sum_{J,\pi} (2J+1) \frac{\left<\Gamma_\mathrm{p}^0\right> \left<\Gamma_\gamma \right>}{\left<\Gamma_\gamma \right>+\left<\Gamma_\mathrm{p} \right>+\left<\Gamma_\mathrm{n} \right>+\left<\Gamma_\alpha \right>} \quad,
\end{equation}
where the superscript on a width indicates a partial width leading to a single final state whereas widths without 
superscript denote sums of partial widths including all energetically possible 
transitions in the given channel,
\begin{equation}
\label{eq:totwid}
\left<\Gamma_x \right>=\left<\Gamma_x^0 \right>+\left<\Gamma_x^1 \right>+\left<\Gamma_x^2 \right>+\dots = \sum_m \left<\Gamma_x^m \right> \quad.
\end{equation}
These are transitions connecting the final states $m$ in each channel with a compound (entry) state at a 
given energy and with spin $J$ and parity $\pi$.

The calculation of averaged widths requires certain nuclear properties to be known. For the particle widths 
these are mainly the level scheme of low-lying levels, as transitions to these dominate the total width 
\cite{Rauscher_Sensitivity,sensi}, and the optical potentials used to compute the strength of each transition. 
For $\left<\Gamma_\gamma \right>$,  however, these are the photon strength functions determining the 
strength of each transition and the nuclear level density, as $\gamma$-ray transitions to states with high 
excitation energy dominate, above the region of well resolved, isolated levels \cite{gammaenergies}. A variation 
of any of these properties translates into a variation of the width and the sensitivities $s$ shown in Figs.~\ref{fig:ratesensi} and 
\ref{fig:Sensitivity} make it possible to judge the impact on the rate and cross section.

Equation \eqref{eq:totfrac} holds for the total cross section, whereas for the partial cross sections a slightly modified version applies,
\begin{equation}
\label{eq:partfrac}
\sigma^m \propto \sum_{J,\pi} (2J+1) \frac{\left<\Gamma_\mathrm{p}^0\right> \left<\Gamma_\gamma^m \right>}{\left<\Gamma_\gamma \right>+\left<\Gamma_\mathrm{p} \right>+\left<\Gamma_\mathrm{n} \right>+\left<\Gamma_\alpha \right>} \quad.
\end{equation}
The only difference is found in the numerator where the total $\gamma$ width is replaced by the partial width 
leading to state $m$. This explains the different behavior of the partial cross sections to a variation of the 
nuclear level density in $^{75}$As. A variation of the nuclear level density only affects the total width 
$\left<\Gamma_\gamma \right>$ which sums (or integrates) over all accessible levels, as shown in Eq.~\eqref{eq:totwid}, 
whereas obviously it does not enter the partial width $\left<\Gamma_\gamma^m \right>$. Because the total $\gamma$ width 
appears both in the numerator and denominator of Eq.~(\ref{eq:totfrac}), the sensitivity to a variation of the nuclear 
level density is the same as to a multiplication of the $\gamma$ width with a factor. For the energy range interesting 
for the $\gamma$ process we found no sensitivity to such a variation, as illustrated by Fig.~\ref{fig:Sensitivity}.

\begin{figure}
\includegraphics[width=\columnwidth]{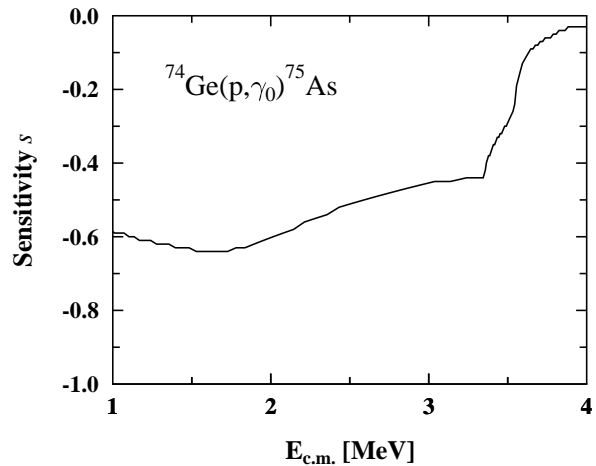}
\caption{\label{fig:nldsensi} Sensitivity $s$ of the $^{74}$Ge(p,$\gamma_0$)$^{75}$As partial cross section when 
varying the nuclear level density in the $\gamma$ channel by a factor of two. Negative sensitivities imply that the 
cross section is changed opposite to the variation of the level density.}
\end{figure}

However, Fig.~\ref{fig:nldsensi} shows the sensitivity of the cross section for $^{74}$Ge(p,$\gamma_0$)$^{75}$As 
as an example for the sensitivities of partial cross sections to a variation of the nuclear level density in $^{75}$As. A nonzero 
sensitivity is found also below the (p,n) threshold, in the astrophysically relevant energy range. The negative values 
shown imply that the cross section changes opposite to the variation factor of the nuclear level density; that is, the 
cross section decreases when the level density is increased and vice versa. This can easily be understood by realizing 
that the total $\gamma$ width, which is modified by a change in the level density, is appearing only in the denominator 
of the width ratio in Eq.~\eqref{eq:partfrac}. If $\left<\Gamma_\gamma \right>$ comprises a large part of the denominator 
(and this is the case as the $\alpha$ and proton widths are very small and we are below the neutron emission threshold), 
the variation cannot cancel with the numerator, as it would be the case for the total cross section. It is important to 
note, however, that a multiplication of both the partial and total $\gamma$ width by the same factor would again lead 
to the same sensitivity as given by Eq.~\eqref{eq:totfrac} and shown in Fig.~\ref{fig:Sensitivity}. This would be 
equivalent to a (energy-independent) variation of the photon strength function.

Total \textit{and} partial cross sections, and their energy dependence, 
are well reproduced by theory when using modified proton widths across 
the measured energy range. Because the sensitivities of the partial cross
section to the $\gamma$ widths change with energy, this leads to the 
conclusion that the $\gamma$ widths are predicted well without need of
further modification. This includes the combination of photon strength 
function and nuclear level density in $^{75}$As entering the $\gamma$ 
width in the total cross sections as well as the nuclear level density
of $^{75}$As at the excitation energies relevant for the $\gamma$-ray 
transitions. 

It was shown that the relevant excitation energies of the final states
of the $\gamma$-ray transitions dominating $\left<\Gamma_\gamma \right>$ 
are located $2-4~\mathrm{MeV}$ below the compound formation energy 
\cite{gammaenergies}. It is a curious fact that these relevant energies 
are very similar for the reactions $^{74}$Ge(p,$\gamma$)$^{75}$As and 
$^{74}$As(n,$\gamma$)$^{75}$As at the same astrophysical plasma temperature 
\cite{Gamow}. Hence, these two reactions, which have the same exit channel,
can be described with the same $\gamma$ widths. In addition, the 
$^{74}$As(n,$\gamma$) rate at $\gamma$ process temperatures is exclusively 
sensitive to the $\gamma$ widths \cite{sensi}. 
%This is because the proton separation energy in $^{75}$As is lower than the neutron separation energy by roughly 
%$3~\mathrm{MeV}$ while the relevant energy window for $^{74}$Ge(p,$\gamma$) is at roughly $3~\mathrm{MeV}$ higher energies than the one for 
%$^{74}$As(n,$\gamma$) \cite{Gamow}. 
%From the present $^{74}$Ge(p,$\gamma$) experiment we can therefore also derive that 
%the $\gamma$ width in $^{74}$As(n,$\gamma$) is well predicted. This is an important finding because the $^{74}$As(n,$\gamma$) 
%rate at $\gamma$-process temperatures is only sensitive to the $\gamma$ width, contrary to any astrophysical charged particle 
%capture reaction \cite{sensi}. 
As a consequence the $^{74}$As(n,$\gamma$) rate is predicted well using the 
standard $\gamma$ widths of the current version of the SMARAGD code.

\begin{table}
\caption{\label{tab:ngrate} Stellar reactivity $N_A \left< \sigma v \right>^*$ and g.s.\ contribution $X$ (taken from \cite{sensi}) 
for $^{74}$As(n,$\gamma$)$^{75}$As as function of plasma temperature.}
\begin{ruledtabular}
\begin{tabular}{c c c c r@{$\times$}l c c c c}
$T$ & & & &\multicolumn{2}{c}{Reactivity} & & & &\textit{X} \\
$[\mathrm{GK}]$& & & & \multicolumn{2}{c}{[cm$^3$s$^{-1}$mole$^{-1}$]} & & & &\\ 
\hline
	&     &    &   &	    &		&     &	    &	  &	\\
  0.10  & $~$ &$~$ &$~$&  1.035$~~$ &  $10^{8}$ & $~$ & $~$ & $~$ &1.00 \\
  0.15  & $~$ &$~$ &$~$&  1.058$~~$ &  $10^{8}$ & $~$ & $~$ & $~$ &1.00 \\
  0.20  & $~$ &$~$ &$~$&  1.088$~~$ &  $10^{8}$ & $~$ & $~$ & $~$ &1.00 \\
  0.30  & $~$ &$~$ &$~$&  1.145$~~$ &  $10^{8}$ & $~$ & $~$ & $~$ &1.00 \\
  0.40  & $~$ &$~$ &$~$&  1.185$~~$ &  $10^{8}$ & $~$ & $~$ & $~$ &0.99 \\
  0.50  & $~$ &$~$ &$~$&  1.198$~~$ &  $10^{8}$ & $~$ & $~$ & $~$ &0.97 \\
  0.60  & $~$ &$~$ &$~$&  1.180$~~$ &  $10^{8}$ & $~$ & $~$ & $~$ &0.93 \\
  0.70  & $~$ &$~$ &$~$&  1.136$~~$ &  $10^{8}$ & $~$ & $~$ & $~$ &0.89 \\
  0.80  & $~$ &$~$ &$~$&  1.076$~~$ &  $10^{8}$ & $~$ & $~$ & $~$ &0.83 \\
  0.90  & $~$ &$~$ &$~$&  1.008$~~$ &  $10^{8}$ & $~$ & $~$ & $~$ &0.78 \\
  1.00  & $~$ &$~$ &$~$&  9.414$~~$ &  $10^{7}$ & $~$ & $~$ & $~$ &0.72 \\
  1.50  & $~$ &$~$ &$~$&  6.853$~~$ &  $10^{7}$ & $~$ & $~$ & $~$ &0.50 \\
  2.00  & $~$ &$~$ &$~$&  5.423$~~$ &  $10^{7}$ & $~$ & $~$ & $~$ &0.36 \\
  2.50  & $~$ &$~$ &$~$&  4.555$~~$ &  $10^{7}$ & $~$ & $~$ & $~$ &0.27 \\
  3.00  & $~$ &$~$ &$~$&  3.965$~~$ &  $10^{7}$ & $~$ & $~$ & $~$ &0.21 \\
  3.50  & $~$ &$~$ &$~$&  3.523$~~$ &  $10^{7}$ & $~$ & $~$ & $~$ &0.17 \\
  4.00  & $~$ &$~$ &$~$&  3.165$~~$ &  $10^{7}$ & $~$ & $~$ & $~$ &0.14 \\
  4.50  & $~$ &$~$ &$~$&  2.855$~~$ &  $10^{7}$ & $~$ & $~$ & $~$ &0.11 \\
  5.00  & $~$ &$~$ &$~$&  2.573$~~$ &  $10^{7}$ & $~$ & $~$ & $~$ &0.10 \\
  6.00  & $~$ &$~$ &$~$&  2.054$~~$ &  $10^{7}$ & $~$ & $~$ & $~$ &0.07 \\
  7.00  & $~$ &$~$ &$~$&  1.581$~~$ &  $10^{7}$ & $~$ & $~$ & $~$ &0.05 \\
  8.00  & $~$ &$~$ &$~$&  1.165$~~$ &  $10^{7}$ & $~$ & $~$ & $~$ &0.04 \\
  9.00  & $~$ &$~$ &$~$&  8.247$~~$ &  $10^{6}$ & $~$ & $~$ & $~$ &0.03 \\
 10.00  & $~$ &$~$ &$~$&  5.650$~~$ &  $10^{6}$ & $~$ & $~$ & $~$ &0.02
\end{tabular}
\end{ruledtabular}
\end{table}

\begin{table}
\caption{\label{tab:ngcoeffs}REACLIB parameters for $^{74}$As(n,$\gamma$) and its reverse reaction, obtained from fitting the reactivities shown in Table~\ref{tab:ngrate}.}
\begin{ruledtabular}
\begin{tabular}{ccc}
Parameter & \multicolumn{1}{c}{(n,$\gamma$)} & \multicolumn{1}{c}{($\gamma$,n)} \\
\hline
       &                         &                       \\
$a_0$  & $6.592306\times 10^{1}$ & $8.983183\times 10^{1}$ \\
$a_1$  &  $-4.470440\times 10^{-1}$ & $-1.193209\times 10^{2}$ \\
$a_2$  & \multicolumn{2}{c}{$3.601952\times 10^{1}$}  \\
$a_3$  &  \multicolumn{2}{c}{$-8.985803\times 10^{1}$}   \\
$a_4$  & \multicolumn{2}{c}{$7.297599$}  \\
$a_5$  & \multicolumn{2}{c}{$-5.682330\times 10^{-1}$}  \\
$a_6$  & $3.457422\times 10^{1}$ & $3.607422\times 10^{1}$
\end{tabular}
\end{ruledtabular}
\end{table}

\begin{figure}
\includegraphics[width=\columnwidth]{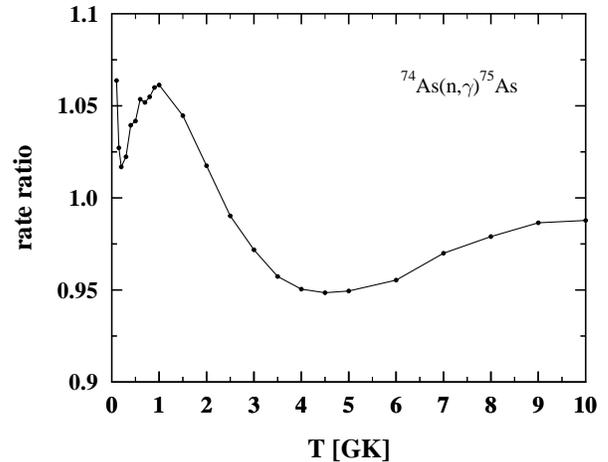}
\caption{\label{fig:ngratio} Ratio of the stellar rate from Table~\ref{tab:ngrate} and the rate given in Ref.~\cite{ADNDT}.}
\end{figure}

As for the $^{74}$Ge(p,$\gamma$)$^{75}$As reaction, we also give the reactivities and REACLIB fit 
coefficients for the reaction $^{74}$As(n,$\gamma$)$^{75}$As in Tables~\ref{tab:ngrate} and \ref{tab:ngcoeffs},
and show a comparison to the rate from Ref.~\cite{ADNDT} in Fig.~\ref{fig:ngratio}. The new rate differs 
only by a few percent from the rate given in Ref.~ \cite{ADNDT} but the information gained from the present 
(p,$\gamma$) measurement puts the rate on a firmer basis.

In the astrophysical $\gamma$ process, the $^{75}$As($\gamma$,n)$^{74}$As rate competes 
with the $^{75}$As($\gamma$,p) and $^{75}$As(p,n) rates and therefore also impacts the production of $^{74}$Se. 
Moreover, when this reaction is well predicted it is also likely that the reaction rate for
 $^{75}$Se($\gamma$,n)$^{74}$Se is predicted equally well.

\section{Summary and conclusion}

The total cross sections of the reaction $^{74}$Ge(p,$\gamma$)$^{75}$As were measured between $2.1~\mathrm{MeV}$ 
and $3.7~\mathrm{MeV}$ using the in-beam $\gamma$ spectroscopy technique with HPGe detectors of high efficiency. 
This energy range covers a considerable fraction of the astrophysically relevant energy range and thus 
the results make it possible to draw conclusions for the prediction of the astrophysical reaction rate used in 
$\gamma$-process simulations. The in-beam technique with high-resolution detectors is a very sensitive 
tool that allows, in addition to the determination of total cross sections, the investigation of 
partial cross sections as well. Twelve of the direct feeding transitions were observed and investigated.

The astrophysical conclusions drawn are important for the synthesis of the lightest $p$ nuclide, $^{74}$Se. 
Using the present data, it was possible to improve the predictions of cross sections and also of the 
astrophysical reaction rate for $^{74}$Ge(p,$\gamma$)$^{75}$As (and its reverse reaction). A renormalization 
of the proton width of the $^{74}$Ge(p,$\gamma$) reaction by a factor of two was required to achieve a good reproduction of the magnitude 
of the total cross sections and their energy dependence. The same renormalization can consistently 
reproduce the partial cross sections of this reaction. This factor of two is probably attributable to deficiencies in the p+$^{74}$Ge optical potential,
which calls for 
further investigations and will be important to improve the global optical potential. Nevertheless, 
the derived astrophysical reaction rate is higher by only about 28\% than the previous prediction by Ref.~\cite{ADNDT}.

The combination of partial and total cross sections allowed further conclusions which were specific for this case. 
The combined data allowed to test the prediction of the $^{74}$As(n,$\gamma$)$^{75}$As cross 
sections at the astrophysically relevant energies. This allowed to put the prediction of this reaction (and its reverse) 
on a firm grounding. The newly calculated rate, however, differs only by a few percent 
from the rate standard given in Ref.~\cite{ADNDT}.

Finally, our calculations were sensitive enough to check 
the spin and parity assignments of the excited states 
appearing as final states in the partial cross sections.
To describe the data, it was necessary to use a 
3/2$^-$, 5/2$^-$ doublet at 1075 keV, contrary to the 
3/2$^-$ assignment in Ref.~\cite{ENSDFnew}. A change in parity 
from 5/2$^+$ to 5/2$^-$ was required for the final state 
at 401 keV.

The above results underline the power of the in-beam technique
with high-efficiency detectors, combined with a sensitivity 
analysis of the cross sections, as a tool for nuclear astrophysics 
as well as nuclear structure investigations.

\begin{acknowledgments}
The authors acknowledge the help of the accelerator staff of Demokritos. Moreover, we thank K.-O. 
Zell for the target preparation. A.S. is member of the Bonn-Cologne Graduate School of Physics and 
Astronomy. This project has been supported by the Deutsche Forschungsgemeinschaft under the 
contract ZI 510/5-1 and partly by the FP7 REGPOT/LIBRA project (Grant No. 230123). T.R. is supported 
by the European Commission within the FP7 ENSAR/THEXO project and by the EuroGENESIS program.
%Support von SH, LIBRA? Wenn wir TALYS nutzen, hier Vivian fuer Diskussionen danken.
\end{acknowledgments}

\end{document}